\documentclass[aps,twocolumn,showpacs,superscriptaddress, preprintnumbers]{revtex4}
\usepackage{amsfonts}
\usepackage{amsmath}
\usepackage{hyperref}
\usepackage{scalefnt}

\newtheorem{theorem}{Theorem}

\newenvironment{proof}[1][Proof]{\noindent\textbf{#1.} }{\ \rule{0.5em}{0.5em}}

\DeclareMathOperator{\Tr}{Tr}
\input{tcilatex}

\begin{document}

\title{The disentangling power of unitaries}
\author{Lieven Clarisse}
\email{lc181@york.ac.uk}
\affiliation{Dept. of Mathematics, The University of York, Heslington, York YO10 5DD, U.K.}
\author{Sibasish Ghosh}
\email{sibasish@imsc.res.in}
\affiliation{The Institute of Mathematical Sciences, C.I.T Campus, Taramani, Chennai 600
113, India}
\author{Simone Severini}
\email{ss54@york.ac.uk}
\affiliation{Dept. of Mathematics, The University of York, Heslington, York YO10 5DD, U.K.}
\affiliation{Dept. of Computer Science, The University of York, Heslington, York YO10
5DD, U.K.}
\author{Anthony Sudbery}
\email{as2@york.ac.uk}
\affiliation{Dept. of Mathematics, The University of York, Heslington, York YO10 5DD, U.K.}

\begin{abstract}
We define the disentangling power of a unitary operator in a similar way as
the entangling power defined by Zanardi, Zalka and Faoro [PRA, 62, 030301].
A general formula is derived and it is shown that both quantities are
directly proportional. All results concerning the entangling power can
simply be translated into similar statements for the disentangling power. In
particular, the disentangling power is maximal for certain permutations
derived from orthogonal latin squares. These permutations can therefore be
interpreted as those that distort entanglement in a maximal way.
\end{abstract}

\maketitle

\section{Introduction}

The dynamics of a closed quantum system can be described by a unitary
operator. In the bipartite scenario, it is natural to study these operators
with respect to their entanglement generating and breaking abilities. In
particular one can study the so-called entangling and disentangling power of
unitaries acting on pure states. These attempt to quantify the increase or
decrease of the entanglement of a bipartite state under a unitary operation.
There are several ways of defining the (dis)entangling power. We follow the
approach of Ref.~\cite{ZZF00} (for an alternative approach see Ref.~\cite%
{LSW05} and references therein). In Ref.~\cite{CGSS05} we studied the
entangling power of permutations, in particular we gave a complete
classification of the permutations maximizing the entangling power. It turns
out that the maximum value over all permutations is the maximum that can be
attained over all unitaries, with possible exception for $6\otimes 6$
systems. In this note, we derive a general formula for the disentangling
power of permutations, which turns out to be proportional to the entangling
power. Thus the permutations with highest entangling power, also have
highest disentangling power and therefore change entanglement in a maximal
way.

\section{Entangling power of unitaries}

In this section we briefly review the definitions and results of Ref.~\cite%
{ZZF00, Zanardi00} in the notations of Ref.~\cite{CGSS05}.

Let $\mathcal{H}_{A}$, $\mathcal{H}_{B}$ and $\mathcal{H}=\mathcal{H}%
_{A}\otimes \mathcal{H}_{B}$ be Hilbert spaces where $\dim \mathcal{H}%
_{A}=\dim \mathcal{H}_{B}=d$. As pure state entanglement measure we use the
normalized linear entropy $S_{L}(\cdot )$ of the reduced density matrix. It
is defined as 
\begin{equation*}
S_{L}(|\psi \rangle ):=\displaystyle{\frac{d}{d-1}(1-\Tr\rho ^{2})},\quad %
\mbox{where}\quad \rho =\Tr_{B}|\psi \rangle \langle \psi |,
\end{equation*}%
for $|\psi \rangle \in \mathcal{H}$. We define the entangling power $%
\epsilon (U)$ of a unitary $U\in \mathcal{U(}\mathcal{H})\cong U(d^{2})$ as
the average amount of entanglement produced by $U$ acting on a distribution
of product states: 
\begin{equation}
\epsilon (U):=\int_{\langle \psi _{1}|\psi _{1}\rangle =1}\int_{\langle \psi
_{2}|\psi _{2}\rangle =1}S_{L}(U|\psi _{1}\rangle |\psi _{2}\rangle )d\psi
_{1}d\psi _{2},  \label{epu}
\end{equation}%
where $d\psi _{1}$ and $d\psi _{2}$ are normalized probability measures on
unit spheres.

With each operator $X$ on $\mathcal{H}\cong \mathcal{H}_{1}\otimes \mathcal{H%
}_{3}$ we can associate a state vector $|X\rangle $ in $\mathcal{H}\otimes 
\mathcal{H}\cong \mathcal{H}_{1}\otimes \mathcal{H}_{2}\otimes \mathcal{H}%
_{3}\otimes \mathcal{H}_{4}$ as 
\begin{equation}
|X\rangle _{A|B}=|X\rangle _{12|34}:=(X_{13}%
\otimes I_{24})|\Psi _{+}\rangle _{13|24},  \label{vecfrommat}
\end{equation}%
where 
\begin{equation*}
|\Psi _{+}\rangle _{13|24}=\frac{1}{d}\sum_{i,j=1}^{d}|ij\rangle
_{13}\otimes |ij\rangle _{24}
\end{equation*}%
and $I$ stands for the identity operator. It easily follows that $%
S_{L}(|I\rangle )=S_{L}(|S\rangle )=1$, with $S=\sum_{ij}^{d}|ij\rangle
\langle ji|$ the swap operator. This isomorphism allows us to rewrite
equation (\ref{epu}) in a form that doesn't require averaging, as in the
following theorem.

\begin{theorem}[Zanardi \protect\cite{Zanardi00}]
\label{zana} The entangling power of a unitary $U\in \mathcal{U(}\mathcal{H}%
) $ is given by 
\begin{equation}
\epsilon (U)=\frac{d}{d+1}[S_{L}(|U\rangle )+S_{L}(|US\rangle
)-S_{L}(|S\rangle )].
\end{equation}%
It follows that $0\leq \epsilon (U)\leq \frac{d}{d+1}$.
\end{theorem}

The maximum $\epsilon(U)= {d}/(d+1)$ is reached for special permutations
(except for $d=2,6$) constructed from orthogonal latin squares, see Ref.~%
\cite{CGSS05}.

\section{Disentangling power of unitaries}

With this we define the \emph{disentangling power} of a unitary $U\in 
\mathcal{U(}\mathcal{H})\cong U(d^{2})$ as 
\begin{equation}
\delta (U):=1-\int_{V\in \mathcal{U}}\int_{W\in \mathcal{U}}S_{L}(U(V\otimes
W)|\psi _{+}\rangle )dVdW,  \label{dpu}
\end{equation}%
where $V,W\in U(d)$, $dV,dW$ are the Haar measure on $U(d)$ and $|\psi
_{+}\rangle =\frac{1}{\sqrt{d}}\sum_{i=1}^{d}|ii\rangle $. Thus, the
disentangling power of a unitary is defined as the average decrease of the
entanglement of the states obtained by applying the unitary on random
maximally entangled states. Note that we could have chosen $V=I$, but for
what follows, the above form is easier to work with.

Following a similar strategy as in \cite{ZZF00, Zanardi00} we now present
the analogue of Theorem~1 for the disentangling power.

\begin{theorem}
The disentangling power of a unitary $U\in \mathcal{U}(\mathcal{H})$ is
given by 
\begin{equation}
\delta(U)=\frac{1}{d-1}[S_{L}(|U\rangle) +
S_{L}(|US\rangle)-S_{L}(|S\rangle)].  \label{newe}
\end{equation}
\end{theorem}

\begin{proof}
(sketch) In a first step, we can rewrite Equation~\ref{dpu} in a similar
form as Equation~(3) from Ref.~\cite{ZZF00}. The method of doing so is
completely analogous; one obtains 
\begin{align}
\delta (U)& =\frac{1}{d-1}\left[ d\Tr((U_{12}\otimes U_{34})\Omega
(U_{12}\otimes U_{34})^{\dagger }\right.  \notag \\
& \left. (S_{13}\otimes I_{24}))-1\right] ,  \label{duuu}
\end{align}%
with 
\begin{align}
\Omega & =\int_{V,W\in \mathcal{U}}(V_{1}\otimes V_{3}\otimes W_{2}\otimes
W_{4})P_{13|24}^{+}  \notag \\
& (V_{1}\otimes V_{3}\otimes W_{2}\otimes W_{4})^{\dagger }dVdW.
\end{align}%
Here, we have introduced four subsystems, and subscripts denote on which
subsystem the operators act. We used $P_{13|24}^{+}$ to denote the maximally
entangled state between subsystems 13 and 24. Integrals of this form can be
evaluated using the fact that $V\otimes V$-invariant operators are linear
combinations of $I$ and $S$, see Ref.\ \cite{VW01}. This particular integral
was evaluated as (see Equation~(27) in Ref.\ \cite{VW01}) 
\begin{align}
\Omega & =\frac{2}{d^{3}(d^{2}-1)}[(d-1)P_{13}^{+}\otimes
P_{24}^{+}+(d+1)P_{13}^{-}\otimes P_{24}^{-}]  \notag \\
& =\frac{1}{d^{2}(d^{2}-1)}[I_{13}\otimes I_{24}+S_{13}\otimes S_{24}] 
\notag \\
& -\frac{1}{d^{3}(d^{2}-1)}[S_{13}\otimes I_{24}+I_{13}\otimes S_{24}].
\label{oooo}
\end{align}%
According to Equation~6 in Ref.\ \cite{Zanardi00} we have 
\begin{align}
S_{L}(|U\rangle )& =\frac{d^{2}}{d^{2}-1}\bigl[1-\frac{1}{d^{4}}\Tr%
((U_{12}\otimes U_{34})\cdot  \notag \\
& (S_{13}\otimes I_{24})\cdot (U_{12}\otimes U_{34})^{\dagger }S_{13}\otimes
I_{24})\bigr].
\end{align}

Substituting Equation~(\ref{oooo}) in Equation~(\ref{duuu}) and using the
above expression for $S_{L}(|U\rangle )$ one obtains readily Equation~(\ref%
{newe}).
\end{proof}

From this theorem follows that 
\begin{equation}
\delta(U)=\frac{d+1}{d(d-1)}\epsilon(U),
\end{equation}
so that the entangling power is proportional to the disentangling power.
With this in mind, all results for the entangling power can simply be
translated into statements of the disentangling power. For instance we have
the following analogue of Theorem~4 and its Corollary from Ref.~\cite{CGSS05}%
. \bigskip

\begin{theorem}
The maximum value of the disentangling power $\delta(U)$ over all unitaries
is achieved for the unitaries with maximum entangling power. For $d\neq 2,6$
this maximum value is given by 
\begin{align}
\delta(U)=\frac{1}{d-1},
\end{align}
and can be attained by permutation matrices only.
\end{theorem}

We would like to thank Dan Browne for some helpful comments on the manuscript.

\end{document}